\documentclass[11pt,a4paper]{article}
\usepackage[T1]{fontenc}
\usepackage[utf8]{inputenc}
\usepackage{lmodern}
\usepackage[margin=1in]{geometry}
\usepackage{amsmath,amssymb,amsthm,mathtools}
\usepackage{microtype}
\usepackage{booktabs,array}
\usepackage{tikz}
\usetikzlibrary{arrows.meta,positioning,fit}
\usepackage[colorlinks=true,linkcolor=blue!55!black,citecolor=blue!55!black,urlcolor=blue!55!black]{hyperref}
\newcommand{\ket}[1]{\lvert #1\rangle}
\newcommand{\bra}[1]{\langle #1\rvert}
\newcommand{\braket}[2]{\langle #1\vert #2\rangle}
\newcommand{\Tr}{\operatorname{Tr}}
\newcommand{\Pe}{\Pi_E}
\newcommand{\PiI}{\Pi_I}
\newcommand{\heff}{\widehat{\hbar}}
\title{\textbf{Leggett--Garg Tests of Collective\\
Excitation--Inhibition Dynamics in Neuronal Circuits}\\[0.5em]
\large Coherent E--I dynamics, negative-result measurements,\\
and experimental challenges}
\author{Partha Ghose\thanks{Email: partha.ghose@gmail.com}\\Tagore Centre for Natural Sciences and Philosophy,\\
Rabindra Tirtha, New Town, Kolkata 700156, India}
\date{}
\begin{document}
\maketitle

\begin{abstract}
Recent approaches to quantum-like signatures in brain and cognition include Bell-type consistency tests, temporal inequalities, and Schr\"odinger-inspired descriptions of collective brain dynamics. We propose a complementary neuronal test based on Leggett--Garg (LG) temporal correlations: could a suitably defined collective excitation--inhibition degree of freedom exhibit the correlations predicted by coherent two-state Schr\"odinger evolution? Two orthogonal reference configurations, \(\ket{E}\) and \(\ket{I}\), span a hypothetical collective state space with Hamiltonian \(H_S=\heff\Omega\sigma_x/2\). Two complementary feedforward synaptic detector pathways are represented by projector-selective interactions: an EPSP-producing detector couples only to \(\ket{E}\), and an IPSP-producing detector only to \(\ket{I}\). An ideal null result then identifies the complementary configuration without detector-induced evolution on that branch. We construct an explicit four-output-cell model whose local activity operators restrict to the desired projectors, specify the corresponding detector unitary operations, and obtain \(K_3(\tau)=2\cos(\Omega\tau)-\cos(2\Omega\tau)\), with maximum \(3/2\). If the spatially distinct excitatory and inhibitory complexes are treated as separate subsystems, a coherent superposition of the two alternative collective configurations is entangled across that partition, even without spatial overlap. The test concerns temporal correlations and makes no appeal to Bell nonlocality. We extend the ideal model to calibrated detector efficiency, false-positive events, imperfect null results, timing jitter and disturbance-adjusted LG bounds. Finally, we set out a concrete experimental agenda, inviting neuroscientists to address the central challenges of preparing, controlling and measuring the proposed collective degrees of freedom. 

We also establish the structural correspondence between
four-time Leggett--Garg and Bell--CHSH correlations,
using the Choi--Jamio{\l}kowski channel--state duality.
The resulting four-time inequality yields the Tsirelson
bound $K_4^{\max}=2\sqrt{2}$ for the ideal E--I model.
\end{abstract}

\section{Motivation and scope}

The search for experimentally accessible quantum-like structure in brain and cognitive dynamics is developing along several complementary lines. Quantum-probabilistic models of cognition, reviewed by Pothos and Busemeyer \cite{PothosBusemeyer2022}, have motivated tests of interference, contextuality and the consequences of sequential questions. Waddup, Yearsley, Blasiak and Pothos have examined a temporal Bell inequality of Leggett--Garg form in human memory, reporting suggestive evidence of a violation in their second behavioural experiment \cite{Waddup2023}. At the level of distributed brain activity, Deco, Sanz Perl and Kringelbach introduced complex harmonics decomposition (CHARM), employing a Schr\"odinger-inspired complex representation to reveal low-dimensional organization in whole-brain dynamics \cite{Deco2025}. A complementary information-theoretic proposal by Kominis and collaborators uses Bell-type consistency constraints on latent representations in autoencoders as a route to probing possible nonclassical structure in neural information processing \cite{Kominis2026}.

A central issue for Bell-inspired investigations is the presence of signalling. In the conventional spatial Bell scenario, measurement choices at the two wings are taken not to influence the distant local statistics. Interconnected neural systems, by contrast, exchange signals through multiple pathways. Broom, Naccache, Pothos, Gallus and Blasiak have recently developed a general framework for measuring the local fraction of Bell-type correlations \emph{in the presence of signalling} \cite{Broom2026}. By maximizing the contribution admitting a local explanation, their convex-decomposition method extends Bell analysis beyond the usual non-signalling regime. This development illustrates the increasing attention being paid to operational tests of quantum-like correlations in systems where direct interactions are difficult to eliminate.

We pursue a complementary possibility: \emph{a temporal rather than spatial inequality}. Leggett--Garg (LG) tests \cite{LeggettGarg1985,Emary2014} compare correlations of the same dichotomic variable at different times, using distinct pairwise measurement protocols. They therefore do not require two spatially isolated measurement wings. The central experimental issue becomes the implementation and calibration of the sequential measurements, which we address with a selective relay--detector model. Our proposal is specifically neurophysiological: rather than testing human judgments or abstract latent representations, it concerns two distinguishable collective excitation--inhibition configurations of a neuronal assembly.

Suppose these collective configurations admit coherent amplitude evolution. We ask whether complementary excitatory and inhibitory output pathways could implement negative-result measurements of their collective state and reveal the predicted LG violation. The construction accommodates spatially separated E and I complexes through a joint state space for their internal degrees of freedom; Bell nonlocality is not invoked. We develop the coherent two-state dynamics, an explicit feedforward measurement scheme and the ideal prediction, then extend the analysis to detector inefficiency and disturbance. The closing Discussion identifies the concrete challenges that experimental neuroscientists would need to address in realizing such a test.

A further structural connection links the four-time Leggett--Garg
inequality to the spatial Bell--CHSH inequality. For four times
$t_1<t_2<t_3<t_4$, the temporal combination
\begin{equation}
 K_4=C_{12}+C_{23}+C_{34}-C_{14}\leq2
 \label{eq:four_time_LG}
\end{equation}
has the CHSH algebraic form. Fritz showed that, for the temporal
CHSH scenario with projective measurements, the set of attainable
quantum correlators coincides with that in the spatial CHSH scenario
\cite{Fritz2010}; both have the Tsirelson bound $2\sqrt{2}$.
A Choi--Jamio{\l}kowski channel--state construction provides a
more explicit relation for specified preparations and sequential
projective measurements (Appendix~\ref{app:chsh_lg}). The
four-time expression has the CHSH algebraic form, while its
fixed-observable LG implementation remains a temporal experiment
with separately realized measurement contexts. The locality and clumsiness
loopholes are compared in Appendix~\ref{app:loopholes}.

\section{A collective two-state model}

Let \(S\) denote a central neuronal assembly containing excitatory principal neurons and inhibitory interneurons. Its two orthogonal, hypothetical reference states are
\begin{equation}
 \ket{E},\qquad \ket{I},\qquad \braket{E}{I}=0.
 \label{eq:basis}
\end{equation}
The labels \emph{excitation dominated} and \emph{inhibition dominated} refer to two distinguished configurations of the collective system, each of which may contain both excitatory and inhibitory activity.

The working hypothesis is that the assembly admits an effective two-dimensional state space
\begin{equation}
 \mathcal H_{\mathrm{EI}}=\operatorname{span}\{\ket E,\ket I\},\qquad
 \ket{\psi(t)}=a(t)\ket E+b(t)\ket I,
 \label{eq:superposition}
\end{equation}
with \(\lvert a\rvert^2+\lvert b\rvert^2=1\) for a pure state. This is our central dynamical hypothesis: the two collective configurations form an effective coherent sector with a definite relative phase. Its possible realization for spatially separated complexes is developed below. For the minimal model take
\begin{equation}
 H_S=\frac{\heff\Omega}{2}\sigma_x
 =\frac{\heff\Omega}{2}
 \bigl(\ket E\bra I+\ket I\bra E\bigr),\qquad
 i\heff\frac{d\ket\psi}{dt}=H_S\ket\psi.
 \label{eq:HS}
\end{equation}
Here \(\Omega\) is an effective transition frequency, and \(\heff\) is the action scale used to write the effective evolution law; no numerical value or microscopic identification is required for the LG prediction. The measured dichotomic observable is
\begin{equation}
 Q=\Pe-\PiI=\sigma_z,\qquad
 \Pe=\ket E\bra E,\quad \PiI=\ket I\bra I.
 \label{eq:Q}
\end{equation}
Thus an ideal measurement returns \(q=+1\) for \(E\) or \(q=-1\) for \(I\). Experimentally, a collective activity contrast can serve to select candidate reference configurations and guide the design of the measurement instrument.

The ideal propagator is
\begin{equation}
 U_S(\Delta t)=\exp\!\left(-\frac{i H_S\Delta t}{\heff}\right)
 =\cos\!\left(\frac{\Omega\Delta t}{2}\right)\mathbb I
 -i\sin\!\left(\frac{\Omega\Delta t}{2}\right)\sigma_x.
 \label{eq:propagator}
\end{equation}
A useful feature of this choice of Hamiltonian and observable is that the ideal LG correlators derived below are independent of the initial density operator \(\rho_0\), including its purity.

\subsection{Spatial separation, entangled configurations, and coherence}

The superposition in Eq.~\eqref{eq:superposition} concerns amplitudes of collective configurations. Spatial overlap of the E and I complexes is unnecessary: the relevant degrees of freedom are internal, and their coherent alternatives may be embedded in the joint state space of two spatially separated subsystems. The physical conditions are a well-defined relative phase, a mechanism for creating and preserving coherence, and an interaction capable of mixing the alternatives.

In the present circuit, it is natural to treat the spatially distinct
E and I complexes as subsystems with respective spaces $\mathcal H_E$
and $\mathcal H_I$.  Their joint space is
$\mathcal H_E\otimes\mathcal H_I$.  Let $\ket e_E$ and $\ket0_E$
denote distinguishable active and inactive E configurations, with
analogous states $\ket i_I$ and $\ket0_I$ for I.  The two alternatives
in Eq.~\eqref{eq:basis} can then be realized as product configurations,
\begin{equation}
 \ket E\equiv\ket e_E\otimes\ket0_I,\qquad
 \ket I\equiv\ket0_E\otimes\ket i_I,
 \label{eq:ei_product_basis}
\end{equation}
where $\braket{e_E}{0_E}=\braket{0_I}{i_I}=0$.  Equation~\eqref{eq:superposition}
now takes the explicit form
\begin{equation}
 \ket\Psi
 =a\ket e_E\ket0_I+b\ket0_E\ket i_I.
 \label{eq:ei_entangled}
\end{equation}
For $ab\neq0$ this pure state has Schmidt rank two, and is therefore
\emph{entangled with respect to the E--I subsystem partition}.  This E--I entanglement concerns internal collective degrees of freedom rather than spatially overlapping wave packets. It can contribute to nonclassical temporal correlations in a Leggett--Garg experiment, whose observables are sampled at different times. The later four-output-cell construction, Eq.~\eqref{eq:microstates}, gives the corresponding circuit-level partition.

The coherent state~\eqref{eq:ei_entangled} must be distinguished from
the incoherent mixture
\begin{equation}
 \rho_{\mathrm{mix}}
 =p_E\ket E\bra E+p_I\ket I\bra I,
 \qquad p_E+p_I=1.
 \label{eq:ei_mixture}
\end{equation}
The pure superposition has the phase-sensitive cross term $a b^*\ket E\bra I+\mathrm{h.c.}$ in its density operator, in contrast to the diagonal mixture. Coherence is characterized by $\bra E\rho\ket I\neq0$ in this basis. Phase-sensitive preparation and readout would make this distinction experimentally accessible. The general effective coupling between the alternatives can be written
\begin{equation}
 H_{\mathrm{EI}}=
 \begin{pmatrix}
 \varepsilon_E & g\\
 g^* & \varepsilon_I
 \end{pmatrix}_{\!E,I}
 \label{eq:ei_general_hamiltonian}
\end{equation}
This Hamiltonian mixes the configurations of the spatially separated complexes. The minimal Hamiltonian~\eqref{eq:HS} is the resonant special case after a phase choice.

With $[H_S,Q]\neq0$, coherent transitions between the E--I configurations generate the pairwise temporal correlations derived below. The relevant test is a Leggett--Garg experiment on the collective observable $Q=\Pe-\PiI$, with the time dependence determined by the Hamiltonian and the sequential measurement protocol. This interpretation applies whether $\mathcal H_{\mathrm{EI}}$ is used directly or embedded in $\mathcal H_E\otimes\mathcal H_I$.

\section{An ideal feedforward circuit and its physiological readouts}

\subsection{Architecture}

We distinguish the \emph{central assembly} from the downstream measurement pathway. The assembly has two classes of output neurons, while two separate detector branches terminate in recording cells:
\begin{itemize}
 \item Excitatory principal-cell outputs converge through glutamatergic contacts onto an excitatory readout \(D_E\); its pointer event is an EPSP onset.
 \item Inhibitory interneuron outputs converge through GABAergic contacts onto an inhibitory readout \(D_I\); its pointer event is an IPSP onset (with the detector membrane conditions chosen so that the response is genuinely hyperpolarizing).
 \item Only one detector type is interrogated in a given negative-result run. The readouts have no designed recurrent projections into the central assembly.
\end{itemize}

\begin{figure}[htbp]
\centering
\begin{tikzpicture}[>=Latex, node distance=11mm and 12mm, font=\small,
  cell/.style={draw,rounded corners,minimum width=13mm,minimum height=8mm,align=center},
  detect/.style={draw,rounded corners,minimum width=36mm,minimum height=10mm,align=center}]
 \node[cell,fill=green!9] (e1) at (0,0) {\(e_1\)};
 \node[cell,fill=green!9] (e2) at (1.7,0) {\(e_2\)};
 \node[cell,fill=red!8] (i1) at (5.2,0) {\(i_1\)};
 \node[cell,fill=red!8] (i2) at (6.9,0) {\(i_2\)};
 \node[draw,dashed,rounded corners,fit=(e1)(e2)(i1)(i2),inner sep=6mm,
       label=above:{Central assembly \(S\)}] {};
 \node[detect,fill=green!7] (de) at (0.85,-2.0) {\(D_E\): EPSP onset};
 \node[detect,fill=red!6] (di) at (6.05,-2.0) {\(D_I\): IPSP onset};
 \draw[->] (e1.south) -- (de.north west);
 \draw[->] (e2.south) -- (de.north east);
 \draw[->] (i1.south) -- (di.north west);
 \draw[->] (i2.south) -- (di.north east);
\end{tikzpicture}
\caption{Convergent feedforward readout of the collective E--I system. Excitatory and inhibitory output cells project to separate downstream detector cells. Their controlled coupling is modelled in the following section.}
\label{fig:circuit}
\end{figure}

EPSP and IPSP onsets provide alternative downstream pointer signals associated with the collective configurations. A synaptic delay separates interaction with \(S\) from the visible pointer onset; the interrogation time is defined by the earlier coupling event. Timing dispersion and readout errors are treated in the realistic-detector section.

\subsection{An explicit four-output-cell model}

To examine projector selectivity mathematically, consider two excitatory output cells \(e_1,e_2\) and two inhibitory output cells \(i_1,i_2\). Give each an \emph{idealized binary activity degree of freedom} \(\ket 0,\ket 1\) and corresponding occupation operator \(n\), with \(n\ket0=0\) and \(n\ket1=\ket1\). Define
\begin{equation}
 \ket E=\ket{1,1;0,0},\qquad
 \ket I=\ket{0,0;1,1}.
 \label{eq:microstates}
\end{equation}
The semicolon separates the excitatory and inhibitory output cells. The binary states describe the collective activity configurations within the ideal interrogation model.

Let the local output operators add coherently at the \emph{effective interaction level}:
\begin{equation}
 F_E=\frac{g_E}{2}\left(n_{e_1}+n_{e_2}\right),\qquad
 F_I=\frac{g_I}{2}\left(n_{i_1}+n_{i_2}\right).
 \label{eq:localF}
\end{equation}
They satisfy the exact eigenvalue conditions
\begin{align}
 F_E\ket E&=g_E\ket E,\qquad F_E\ket I=0,\label{eq:FE}\\
 F_I\ket E&=0,\qquad F_I\ket I=g_I\ket I.\label{eq:FI}
\end{align}
Writing \(P=\Pe+\PiI\), we obtain
\begin{equation}
 PF_EP=g_E\Pe,\qquad PF_IP=g_I\PiI.
 \label{eq:restriction}
\end{equation}
In this particular construction the operators are diagonal in the four-cell occupation basis, and the subspace has no leakage under these couplings:
\begin{equation}
 (\mathbb I-P)F_EP=(\mathbb I-P)F_IP=0.
 \label{eq:noleak}
\end{equation}
Hence, \emph{restricted to the proposed two-state subspace}, convergent local operators can reproduce the desired ideal projectors.

Thus, convergent local activity operators give precisely the desired selective coupling on the E--I sector. Their implementation as effective neuronal observables is a central experimental task considered in the Discussion.

\subsection{General selectivity conditions}

For an arbitrary Hermitian convergent coupling \(F_E\), a convenient matrix representation in the collective subspace is
\begin{equation}
 PF_EP=\begin{pmatrix}g_E&\epsilon\\ \epsilon^*&\eta\end{pmatrix}_{\!E,I}.
 \label{eq:generalF}
\end{equation}
The ideal measurement requires \(\epsilon=\eta=0\), together with absence of leakage, \((\mathbb I-P)F_EP=0\). Thus
\begin{equation}
 F_E\ket E=g_E\ket E,\qquad F_E\ket I=0.
 \label{eq:idealFE}
\end{equation}
Analogously, an ideal inhibitory detector must satisfy
\begin{equation}
 F_I\ket E=0,\qquad F_I\ket I=g_I\ket I.
 \label{eq:idealFI}
\end{equation}
These operator conditions express genuine selectivity: on the nominally silent branch, the system--detector interaction acts as the identity. They therefore provide a quantitative target for designing and calibrating the relay pathways.

\section{System--relay--detector measurement dynamics}

\subsection{Ideal controlled coupling}

Use a two-state pointer \(D\), initially \(\ket0_D\), and assume the conditional interaction
\begin{equation}
 H_{SD}^{(E)}=\heff F_E\otimes\sigma_x^{(D)}
 \ \overset{\mathcal H_{\mathrm{EI}}}{=}\ 
 \heff g_E\Pe\otimes\sigma_x^{(D)}.
 \label{eq:Hdet}
\end{equation}
The equality on the right holds on the specified two-dimensional subspace. During a short interrogation pulse, free evolution under \(H_S\) is neglected. Then
\begin{equation}
 U_E(t)=\PiI\otimes\mathbb I_D
 +\Pe\otimes\exp(-ig_Et\sigma_x^{(D)}).
 \label{eq:UE}
\end{equation}
For \(t_E=\pi/(2g_E)\), this gives
\begin{align}
 \ket E\ket0_D&\longmapsto-i\ket E\ket1_D,\label{eq:Eclick}\\
 \ket I\ket0_D&\longmapsto\ket I\ket0_D.\label{eq:Enull}
\end{align}
Accordingly, for \(\ket\psi=a\ket E+b\ket I\),
\begin{equation}
 \ket\psi\ket0_D\longmapsto
 -ia\ket E\ket1_D+b\ket I\ket0_D.
 \label{eq:entangleE}
\end{equation}
A click has probability \(\lvert a\rvert^2\); a null result has probability \(\lvert b\rvert^2\), and the associated conditional assembly state is \(\ket I\). In particular, Eq.~\eqref{eq:Enull} is an \emph{identity action} on the negative-result branch.

With \(H_{SD}^{(I)}=\heff g_I\PiI\otimes\sigma_x^{(D)}\) and \(t_I=\pi/(2g_I)\), the complementary detector obeys
\begin{align}
 \ket E\ket0_D&\longmapsto\ket E\ket0_D,\\
 \ket I\ket0_D&\longmapsto-i\ket I\ket1_D.
\end{align}
A null inhibitory response thus selects \(\ket E\).

\subsection{An explicit intermediate relay}

If one wishes to keep the relay \(R\) separate from the synaptic pointer \(D\), first implement
\begin{equation}
 H_{SR}^{(E)}=\heff g_E\Pe\otimes\sigma_x^{(R)}.
 \label{eq:HSR}
\end{equation}
For pulse area \(g_Et_E=\pi/2\),
\begin{equation}
 (a\ket E+b\ket I)\ket0_R
 \longmapsto -ia\ket E\ket1_R+b\ket I\ket0_R.
\end{equation}
A relay-to-detector exchange coupling
\begin{equation}
 H_{RD}=\heff\kappa
 \left(\sigma_+^{(R)}\sigma_-^{(D)}
       +\sigma_-^{(R)}\sigma_+^{(D)}\right)
 \label{eq:HRD}
\end{equation}
acts for \(\kappa t_D=\pi/2\) as
\begin{equation}
 \ket{1_R,0_D}\longmapsto-i\ket{0_R,1_D},\qquad
 \ket{0_R,0_D}\longmapsto\ket{0_R,0_D}.
\end{equation}
The combined ideal transformation is therefore
\begin{equation}
 (a\ket E+b\ket I)\ket{0_R,0_D}
 \longmapsto
 -a\ket E\ket{0_R,1_D}
 +b\ket I\ket{0_R,0_D}.
 \label{eq:fullmap}
\end{equation}
The known phase on the activated branch is irrelevant to the projective outcome statistics and can be compensated if necessary. The negative branch stays uncoupled. The two controlled pulses furnish an explicit relay-to-pointer realization of the ideal instrument.

\subsection{Measurement operators and conditional states}

Ignoring irrelevant outcome-dependent phases, the effective instrument for an excitatory detector is
\begin{equation}
 M_{E,\mathrm{click}}=\Pe,\qquad
 M_{E,\mathrm{null}}=\PiI.
 \label{eq:ME}
\end{equation}
For the inhibitory detector,
\begin{equation}
 M_{I,\mathrm{click}}=\PiI,\qquad
 M_{I,\mathrm{null}}=\Pe.
 \label{eq:MI}
\end{equation}
Both satisfy \(\sum_rM_r^\dagger M_r=\mathbb I\) on \(\mathcal H_{\mathrm{EI}}\). For an arbitrary initial density matrix \(\rho\), a null result from \(D_E\) occurs with probability \(\Tr(\PiI\rho)\) and gives
\begin{equation}
 \rho\longmapsto
 \frac{\PiI\rho\PiI}{\Tr(\PiI\rho)}.
 \label{eq:nullupdate}
\end{equation}
The corresponding statement with \(\Pe\) holds for a null result from \(D_I\).

In the negative-result construction, the detector leaves the untriggered alternative dynamically untouched; the conditional state update in Eq.~\eqref{eq:nullupdate} describes the information gained from a null outcome. This is the ideal noninvasiveness principle underlying the proposed test \cite{Knee2012,Emary2014}.

\section{The Leggett--Garg test}

\subsection{Macrorealist inequality}

For three times \(t_1<t_2<t_3\), let \(Q_i=Q(t_i)\in\{+1,-1\}\). Under macrorealism per se and noninvasive measurability, the same underlying values can be assigned to all three times, with a consistent joint distribution. For each assignment,
\begin{equation}
 Q_1Q_2+Q_2Q_3-Q_1Q_3\leq1.
\end{equation}
Averaging gives the standard inequality
\begin{equation}
 K_3=C_{12}+C_{23}-C_{13}\leq1,
 \qquad C_{ij}=\langle Q_iQ_j\rangle.
 \label{eq:LGI}
\end{equation}

\subsection{Three independent pairwise contexts}

We envisage repeated, nominally identical preparations divided at random into three measurement contexts: \((t_1,t_2)\), \((t_2,t_3)\), and \((t_1,t_3)\). Each context has its own physically implemented earlier-time measurement interaction, as required for a sequential-measurement test.

In a given context \((t_i,t_j)\), the earlier measurement is reconstructed from two \emph{separate} negative-result subensembles. A null result from \(D_I\) identifies \(q_i=+1\); a null result from \(D_E\) identifies \(q_i=-1\). An ideal final projective measurement of \(Q\) determines \(q_j\). The initial probabilities are obtained from the appropriate null-result frequencies, and the conditional data are combined with their correct weights:
\begin{equation}
 \begin{aligned}
 C_{ij}={}&p_i(+1)\,\langle Q_j\rangle_{\text{null }D_I}\\
          &-p_i(-1)\,\langle Q_j\rangle_{\text{null }D_E}.
 \end{aligned}
 \label{eq:weighted}
\end{equation}
Here \(p_i(+1)=\Tr(\Pe\rho(t_i))\) and \(p_i(-1)=\Tr(\PiI\rho(t_i))\), with \(\rho(t_i)\) the unmeasured state at the earlier interrogation time. The two complementary detector settings must represent the same preparation ensemble and the same ideal dichotomic observable. The null-run data are therefore weighted by their independently determined outcome probabilities.

\subsection{Schr\"odinger prediction}

Suppose an ideal earlier measurement yielded \(q_i=\pm1\). Propagation under Eq.~\eqref{eq:HS} gives
\begin{equation}
 P(q_j\mid q_i)=\frac12\left[
 1+q_iq_j\cos\bigl(\Omega(t_j-t_i)\bigr)\right].
 \label{eq:conditional}
\end{equation}
The sequential correlation therefore is
\begin{equation}
 \begin{aligned}
 C_{ij}
 &=\sum_{q_i,q_j=\pm1}q_iq_j\,p_i(q_i)P(q_j\mid q_i)\\
 &=\cos\bigl(\Omega(t_j-t_i)\bigr).
 \end{aligned}
 \label{eq:correlation}
\end{equation}
The second line is independent of \(p_i(q_i)\), hence of the purity of the initial density matrix, for this Hamiltonian and ideal projective measurement scheme.

For equal intervals \(t_2-t_1=t_3-t_2=\tau\),
\begin{equation}
 C_{12}=C_{23}=\cos(\Omega\tau),\qquad
 C_{13}=\cos(2\Omega\tau).
\end{equation}
Consequently,
\begin{equation}
 \boxed{K_3(\tau)=2\cos(\Omega\tau)-\cos(2\Omega\tau).}
 \label{eq:K3}
\end{equation}
Writing \(x=\cos(\Omega\tau)\) gives \(K_3=1+2x-2x^2\), maximized at \(x=1/2\). Thus
\begin{equation}
 \boxed{\Omega\tau=\frac{\pi}{3},\qquad
 K_3^{\mathrm{max}}=\frac32>1.}
 \label{eq:max}
\end{equation}
The factor \(\heff\) cancels from the dimensionless prediction. An observed violation would be inconsistent with the conjunction of macrorealism and noninvasive measurability under the stipulated comparability and measurement assumptions.

For the chosen $H_S$ and $Q$, this result holds for any initial density operator: the relevant control parameter is the dimensionless interval $\Omega\tau$. In particular, the ideal experiment predicts a violation throughout the range $0<\cos(\Omega\tau)<1$, with maximum at Eq.~\eqref{eq:max}.

\subsection{Four-time LG--CHSH extension}

The same ideal model also gives a direct prediction for the
four-time inequality~\eqref{eq:four_time_LG}. For equally spaced
interrogation times $t_{j+1}-t_j=\tau$, the pairwise projective
measurement protocols yield
\begin{equation}
 \begin{aligned}
 C_{12}=C_{23}=C_{34}&=\cos(\Omega\tau),\\
 C_{14}&=\cos(3\Omega\tau),
 \end{aligned}
\end{equation}
and hence
\begin{equation}
 \boxed{K_4(\tau)=3\cos(\Omega\tau)-\cos(3\Omega\tau).}
 \label{eq:K4_prediction}
\end{equation}
At $\Omega\tau=\pi/4$, this reaches
\begin{equation}
 \boxed{K_4^{\mathrm{max}}=2\sqrt{2}>2.}
 \label{eq:K4_max}
\end{equation}
Thus a four-time test provides a natural extension of the proposed
three-time experiment, with the four correlations measured in their
separately implemented pairwise contexts. A derivation of the structural CHSH correspondence and the precise channel--state mapping is given in Appendix~\ref{app:chsh_lg}.

\section{Requirements on the dynamics during measurement}

Because \([H_S,\Pe]\ne0\), an approximately instantaneous measurement is implemented by a pulse short compared with the collective oscillation period. For the excitatory detector,
\begin{equation}
 t_E=\frac{\pi}{2g_E},\qquad
 \Omega t_E\ll1,
 \label{eq:fastpulse}
\end{equation}
with the analogous requirement for \(D_I\). Between measurements, the detector coupling is switched off and the assembly evolves freely under \(H_S\). A readout latency after the interrogation is not, in itself, a different measurement time if the pointer record is faithfully retained.

The stipulated free Hamiltonian on the four-cell reference configurations can be written
\begin{equation}
 H_S=\frac{\heff\Omega}{2}
 \left(\ket E\bra I+\ket I\bra E\right).
 \label{eq:collectiveH}
\end{equation}
In the idealized four-bit representation, an operator with precisely this action on the two reference states is
\begin{equation}
 \begin{aligned}
 H_S=\frac{\heff\Omega}{2}\bigl(&
 \sigma_-^{(e_1)}\sigma_-^{(e_2)}
 \sigma_+^{(i_1)}\sigma_+^{(i_2)}\\
 &+\sigma_+^{(e_1)}\sigma_+^{(e_2)}
 \sigma_-^{(i_1)}\sigma_-^{(i_2)}\bigr).
 \end{aligned}
 \label{eq:fourbody}
\end{equation}
In the four-cell representation, this exchange transforms one collective E--I configuration into the other and can generate entanglement across the two complexes. Its explicit form also separates the task of realizing coherent collective dynamics from that of engineering selective readout.

\section{Realistic measurements: efficiency, false positives, and disturbance}

The ideal relay--detector scheme admits a direct extension to finite readout efficiency, false positives, timing uncertainty and measurement disturbance. We introduce calibration parameters separately for outcome registration and physical back-action.

\subsection{A calibrated binary readout}

Let $q=\pm1$ denote the ideal outcome of a specified measurement
context and $m=\pm1$ its registered binary result.  For a symmetric
readout with efficiency $\eta$ and false-positive probability $f$,
write
\begin{equation}
 P(m=+1\mid q=+1)=\eta,\qquad
 P(m=+1\mid q=-1)=f.
\end{equation}
Consequently,
\begin{equation}
 \mathbb E[m\mid q]=a+bq,\qquad
 a=\eta+f-1,\qquad b=\eta-f.
 \label{eq:ei_readout_calibration}
\end{equation}
For two readouts whose errors are conditionally independent given the
ideal outcomes and are independent of the subsequent system dynamics, and whose calibration parameters are $a_i,b_i$, the
measured pair correlation is
\begin{equation}
 C_{ij}^{\rm obs}
 =a_i a_j+a_i b_j\langle q_j\rangle
 +b_i a_j\langle q_i\rangle+b_i b_j C_{ij}^{\rm ideal}.
 \label{eq:ei_imperfect_pair}
\end{equation}
Here the means and correlations refer to the \emph{same measurement
protocol}; they need not coincide across protocols if measurement
back-action or preparation differs.  When $b_i b_j\neq0$, calibration
allows the ideal correlation to be estimated from this expression,
with uncertainty propagated from the independently measured $\eta_i$
and $f_i$.  For complementary negative-result measurements, selection and reweighting follow Eq.~\eqref{eq:weighted}, using the calibrated probabilities of null and click events. The contrast $\eta_i-f_i$ controls the statistical stability of any inversion. Physical disturbance is characterized independently in the next subsection.

If the detector has a genuine third outcome $m=\varnothing$ (no
registered result), that outcome must be retained in the raw records.
Complete outcome counts, including no-click events, are retained for every measurement context. This permits direct estimation of state- and context-dependent detection losses and makes null-result reliability experimentally testable.

\subsection{Reliability of a null response}

For the excitatory detector, let $p_E$ and $p_I=1-p_E$ be the
prior probabilities of the two alternatives at interrogation.  Under
the calibrated response model, the probability that a registered
null corresponds to the $I$ alternative is
\begin{equation}
 P(I\mid\mathrm{null}\,D_E)=
 \frac{(1-f)p_I}{(1-\eta)p_E+(1-f)p_I}.
 \label{eq:ei_bayes_null}
\end{equation}
The posterior probability approaches unity as $\eta\to1$ and $f\to0$ for nonzero $p_I$. Exchanging $E$ and $I$ gives the inhibitory detector formula. Independent dynamical controls complement these readout-reliability estimates.

\subsection{Imperfect negative-result measurements}

In the ideal protocol, the relay is coupled selectively to one
alternative, and only trials with no relay response are retained to
infer the other alternative.  In a realistic implementation, a no-click
event may result from (i) the alternative being absent, (ii) failure to
detect its presence, or (iii) a detector malfunction.  Conversely,
false clicks may occur when the coupled alternative is absent.  The
relay may also perturb the circuit even when it does not click.

To make the negative-result argument operational, both complementary
relay settings should be implemented and calibrated.  Each setting
requires independent estimates of detection efficiency, false-click
rate, timing jitter, and the influence of the relay coupling on the
subsequent E--I dynamics.  The experimental protocols must physically switch the selective interaction and predetermine the trial selection rule. The readout is then interpreted using both null-event calibration and the dynamics of the selected branch.

\subsection{Disturbance-corrected Leggett--Garg criterion}

Denote by $C_{ij}^{(ij)}$ the correlation measured in the distinct
pairwise protocol that interrogates times $t_i$ and $t_j$.  The
experimentally assembled quantity is
\begin{equation}
 K_3^{\rm obs}=C_{12}^{(12)}+C_{23}^{(23)}-C_{13}^{(13)}.
\end{equation}
The ideal bound $K_3\leq1$ presumes a common preparation and
non-invasive access to the same underlying dichotomic values.  To
quantify departures from this assumption, suppose independent control
experiments establish bounds $\epsilon_{ij}$ on the absolute changes
between the correlations used in the experiment and those of a
specified non-invasive reference model with a common joint
distribution.  The triangle inequality then gives the conservative
bound
\begin{equation}
 K_3^{\rm obs}\leq 1+
 \epsilon_{12}+\epsilon_{23}+\epsilon_{13}.
 \label{eq:ei_disturbance_bound}
\end{equation}
The quantities $\epsilon_{ij}$ are established by independent controls and explicit physical assumptions, with calibration uncertainties, preparation drift and confidence intervals propagated to the corrected bound.

\subsection{Finite timing resolution and loss of contrast}

If nominal measurement times fluctuate, the observed correlations are
averages over the actual time separations.  For the illustrative ideal
coherent model $C(\tau)=\cos(\Omega\tau)$, zero-mean Gaussian jitter
in the \emph{time difference}, of standard deviation $\sigma_\tau$,
yields
\begin{equation}
 C_{\rm jitter}(\tau)
 =\exp\!\left(-\frac{\Omega^2\sigma_\tau^2}{2}\right)
 \cos(\Omega\tau).
\end{equation}
The Gaussian factor quantifies loss of contrast from timing uncertainty. An independently characterized coherence attenuation factor $v(\tau)$ multiplies $\cos(\Omega\tau)$ before application of the detector response model, allowing timing and dynamical effects to be estimated separately.

\subsection{Experimental decision rule}

A credible experiment should report the complete outcome counts for
each pairwise protocol, the calibrated detector confusion matrices,
no-click frequencies, timing distributions, preparation checks, and
independent disturbance controls.  The experimental decision compares a confidence interval for the measured or appropriately calibrated $K_3$ with the independently justified disturbance-adjusted bound. This directly connects measured performance of the detector apparatus with the ideal prediction in Eq.~\eqref{eq:K3}.

\section{Discussion: challenges and opportunities for experimental neuroscience}

The theory specifies an experimental target: a small neuronal assembly with identifiable collective E--I alternatives, controllable transitions between them and complementary detector pathways whose influence on the subsequent dynamics can be measured. The prediction in Eq.~\eqref{eq:K3} supplies a quantitative benchmark, while the realistic-detector analysis provides a framework for approaching it step by step. Translating this construction into neural tissue presents several distinct challenges. We outline them here in the hope that they will stimulate experimental neuroscientists to develop, adapt and test suitable preparations.

\subsection{Identifying and stabilizing a collective E--I sector}

The first challenge is to find a circuit in which two reproducible collective activity configurations can be distinguished with sufficient fidelity. In the four-output-cell illustration, the reference configurations are $\ket{1,1;0,0}$ and $\ket{0,0;1,1}$. Real circuits have many additional degrees of freedom and may occupy intermediate configurations. Simultaneous intracellular recordings, voltage imaging and controlled stimulation could be used to identify candidate E- and I-dominated configurations, quantify transitions out of their proposed two-state sector, and test whether the same preparation can be reproduced across trials. A useful first milestone would be an experimentally calibrated two-state description with measured leakage and preparation variability.

\subsection{Establishing coherent collective dynamics}

The second challenge is to determine whether the two configurations support a coherent relative phase and an effective coupling of the form in Eq.~\eqref{eq:HS}. Spatial separation of the E and I complexes poses no formal obstruction: in the tensor-product description, Eq.~\eqref{eq:ei_entangled} is entangled across their internal degrees of freedom for nonzero amplitudes. What is required physically is controllable preparation and evolution of those alternatives. Neural circuits already exhibit excitation--inhibition rhythms, which provide natural candidates for investigation. Phase-sensitive perturbations, controlled evolution intervals and tests of interference or state reconstruction would be particularly valuable for distinguishing coherent amplitude dynamics from classical rhythmic switching. The four-cell exchange operator in Eq.~\eqref{eq:fourbody} makes the dynamical target explicit; discovering an effective neural mechanism that approximates it is an important theoretical and experimental problem.

\subsection{Engineering selective E and I relay pathways}

The third challenge is to realize and characterize the conditional readout interactions in Eqs.~\eqref{eq:idealFE} and \eqref{eq:idealFI}. An excitatory relay should respond predominantly to the E configuration, while an inhibitory relay should respond predominantly to the I configuration. Selectivity must include the \emph{silent} branch: the uncoupled alternative should experience as little dynamical perturbation as possible. Neural synapses ordinarily transmit through stochastic release, membrane conductances and network interactions, rather than through an established projective quantum instrument. Hence a recorded EPSP or IPSP is a candidate detector signal whose connection to the ideal projector has to be calibrated. Experiments could quantify cross-talk between the two pathways, synaptic failures, response latencies, driving-force dependence of inhibitory signals, and the influence of detector coupling on the central assembly. These measurements would show how closely a biological readout approaches the algebraic selectivity demonstrated by Eq.~\eqref{eq:restriction}.

\subsection{Controlling back-action, timing and statistical errors}

The fourth challenge concerns measurement disturbance. The relevant pairwise protocols require the detector interactions to be physically switched, not merely the recorded data to be selected afterwards. Complementary negative-result measurements should be tested using matched control preparations, including comparisons with the corresponding detector interaction absent. Such controls can bound residual effects on the subsequent E--I evolution and support the disturbance-adjusted inequality in Eq.~\eqref{eq:ei_disturbance_bound}. Independent calibrations of detector efficiency, false-positive events, omitted outcomes, timing jitter and trial-to-trial drift complete the experimental characterization. Repeated preparations and randomized protocol assignment are especially important in a neural environment with common inputs, adaptation and slow changes in excitability.

\subsection{A staged programme and the meaning of a positive result}

An experimentally economical progression would be to (i) identify a stable E--I subspace and characterize its leakage; (ii) seek controllable, phase-sensitive transitions between the alternatives; (iii) construct and calibrate complementary E- and I-selective readouts, including the negative-result branches; and (iv) perform the three independent pairwise LG experiments, with a pre-specified time interval and quantified disturbance. Cultured networks or acute brain slices may offer useful initial settings because stimulation, circuit structure and simultaneous recordings can be controlled more tightly than in an intact brain. Each stage provides informative physiological measurements even if a full LG test remains out of reach.

Finally, the interpretation of a temporal violation should be tied to the controls actually achieved. The ideal expression for $K_3$ is independent of the initial density operator for the stipulated $H_S$, $Q$ and measurement instrument. Thus initial E--I entanglement is neither a separate prerequisite nor a guarantee of a violation: the dynamics, interrogation times and measurement contexts determine the result. A reproducible excess over an independently justified disturbance-adjusted bound would challenge the corresponding macrorealist, noninvasive description. Establishing a specific microscopic entanglement mechanism would remain a separate, phase-sensitive task. The purpose of the proposed scheme is to identify a concrete temporal signature and to encourage experimental investigations capable of deciding whether such collective coherent dynamics can be realized in neuronal circuits.

\section{Conclusion}

We have developed a two-state model of collective excitation--inhibition dynamics, together with an explicit relay--detector construction for temporal Leggett--Garg tests. Coherent superpositions may involve spatially separated E and I complexes through entanglement of their internal configurations. Under the proposed effective Hamiltonian and complementary projective measurements, the ideal prediction is $K_3^{\mathrm{max}}=3/2$, with a four-time extension reaching $K_4^{\mathrm{max}}=2\sqrt{2}$. A realistic extension incorporates finite detection efficiency, false-positive responses, null-result reliability, timing uncertainty and independently calibrated disturbance. These results define a quantitative theoretical benchmark and a sequence of experimentally addressable questions. The channel--state explanation of the related four-time CHSH structure is given in Appendix~\ref{app:chsh_lg}. We hope the identified challenges---from preparation of collective E--I alternatives to state-selective readout and disturbance control---will encourage neuroscientists to examine whether coherent temporal dynamics can be established in suitable neural preparations.

\section*{Appendix: From four-time Leggett--Garg correlations to Bell--CHSH correlations}
\label{app:chsh_lg}

The connection between a four-time Leggett--Garg (LG) test and a
Bell--CHSH test can be understood in three steps. First, their
classical correlation expressions have the same algebraic form.
Second, a simple one-qubit example shows how evolution through a
quantum channel can be represented by a bipartite quantum state.
Finally, the Choi--Jamio{\l}kowski identity and its diagrammatic
``wire-bending'' representation make that connection general
\cite{Fritz2010,Choi1975,Jamiolkowski1972}.

\subsection{The four-time inequality: the elementary connection}
\label{app:fourtime_algebra}

Consider four ordered times $t_1<t_2<t_3<t_4$, with a dichotomic
observable taking values $q_i=\pm1$. For a single assignment of the
four values,
\begin{equation}
\begin{aligned}
 k_4&=q_1q_2+q_2q_3+q_3q_4-q_1q_4\\
    &=q_2(q_1+q_3)+q_4(q_3-q_1)\in\{-2,2\}.
\end{aligned}
\label{eq:appendix_pointwise}
\end{equation}
Thus averaging over a common underlying assignment gives the
classical bound $|K_4|\leq2$. The four correlations in an LG
experiment are, operationally, obtained in separate pairwise
measurement contexts.

The same algebra appears in the spatial CHSH expression
\begin{equation}
 S_{\rm CHSH}=\langle A_0B_0\rangle+\langle A_0B_1\rangle
 +\langle A_1B_0\rangle-\langle A_1B_1\rangle,
 \qquad |S_{\rm CHSH}|\leq2.
 \label{eq:appendix_chsh}
\end{equation}
For instance, setting
\begin{equation}
 A_0=q_2,\quad A_1=q_4,\qquad
 B_0=q_3,\quad B_1=q_1
 \label{eq:appendix_relabel}
\end{equation}
turns $k_4$ into precisely the CHSH combination. This immediately
explains the common classical bound. The next example explains why
a related correspondence also holds for suitably defined \emph{quantum}
correlations.

\subsection{Experimental loopholes: locality in CHSH and clumsiness in LG}
\label{app:loopholes}

The algebraic parallel between CHSH and four-time LG correlations has
an illuminating experimental counterpart: each inequality has a
characteristic loophole reflecting the way its correlations are
measured. In a spatial Bell--CHSH experiment, the \emph{locality
loophole} arises if the choice of measurement or its outcome on one
wing could influence the outcome on the other wing by an ordinary
causal signal. Fast, independently chosen settings and suitable
space-like separation between the relevant setting and outcome events
address this possibility. High detection efficiency is additionally
needed to control the distinct fair-sampling or detection loophole
\cite{Hensen2015}.

For LG tests, the measurement settings are separated primarily in
\emph{time}, not in space. The corresponding concern is the
\emph{clumsiness loophole}: an earlier measurement might perturb the
system and thereby alter a later outcome, even if the system possesses
well-defined classical properties throughout. A measured value
$K_3>1$ or $K_4>2$ would then be compatible with a macrorealist
account that includes this measurement disturbance
\cite{WildeMizel2012}. The central comparison is:

\begin{center}
\small
\renewcommand{\arraystretch}{1.16}
\begin{tabular}{@{}>{\raggedright\arraybackslash}p{1.30in}>{\raggedright\arraybackslash}p{2.20in}>{\raggedright\arraybackslash}p{2.20in}@{}}
\toprule
 & \textbf{Spatial Bell--CHSH} & \textbf{Temporal Leggett--Garg}\\
\midrule
Main risk & Communication between the separated measurement wings & Disturbance of subsequent dynamics by an earlier interrogation\\
Loophole & Locality & Clumsiness (invasiveness)\\
Control strategy & Space-like separation and rapid, independent setting choices & Ideally non-invasive measurements, with independent disturbance controls\\
Neuronal proposal & Spatial isolation is not required by the LG proposal & Complementary E- and I-selective negative-result relay measurements\\
\bottomrule
\end{tabular}
\end{center}

\emph{Ideal negative-result measurement} provides the key LG strategy
\cite{LeggettGarg1985,Knee2012}. A detector is arranged to couple
only to one of two macroscopically distinguishable alternatives. If
it does not register a response, a macrorealist can infer the other
alternative \emph{without an interaction with the occupied branch}.
The complementary detector setting gives the opposite outcome, and
the two null-result ensembles reconstruct the correlation with the
appropriate weights. In our E--I construction these roles are played
by the two projector-selective relays, whose ideal null branches obey
Eqs.~\eqref{eq:idealFE}--\eqref{eq:idealFI} and
\eqref{eq:Enull}.

This strategy addresses the clumsiness loophole in its ideal limit.
For a biological implementation, the crucial experimental task is to
verify that an untriggered relay leaves the subsequent collective
E--I evolution unchanged within a measured tolerance. Missed clicks,
subthreshold synaptic interactions and preparation drift must therefore
be quantified independently; the disturbance-adjusted criterion
\eqref{eq:ei_disturbance_bound} is designed for precisely this purpose.
Thus the structural CHSH--LG correspondence preserves a common
correlation algebra, whereas \emph{locality} and \emph{non-invasiveness}
remain distinct operational requirements.

\subsection{A simple example: representing a NOT gate by a Bell state}
\label{app:not_gate}

Consider the familiar Pauli NOT gate $X$, acting on one qubit:
\begin{equation}
 X\ket0=\ket1,\qquad X\ket1=\ket0,
 \qquad X=\begin{pmatrix}0&1\\1&0\end{pmatrix}.
 \label{eq:appendix_pauli_x}
\end{equation}
This is a \emph{dynamical} operation: an input is transformed into an
output. To give it a \emph{state} representation, introduce a second,
untouched reference qubit and prepare
\begin{equation}
 \ket{\Phi^+}=\frac{\ket{00}+\ket{11}}{\sqrt2}.
 \label{eq:appendix_phi_plus}
\end{equation}
We use the convention that the channel acts on the \emph{first}
qubit and the reference is the second. Applying $X$ to the first
qubit gives
\begin{equation}
 (X\otimes\mathbb I)\ket{\Phi^+}
 =\frac{\ket{10}+\ket{01}}{\sqrt2}
 =\ket{\Psi^+}.
 \label{eq:appendix_psi_plus}
\end{equation}
The resulting two-qubit density operator is
\begin{equation}
 \rho_X=\ket{\Psi^+}\!\bra{\Psi^+}
 =\frac12\begin{pmatrix}
  0&0&0&0\\
  0&1&1&0\\
  0&1&1&0\\
  0&0&0&0
 \end{pmatrix}.
 \label{eq:appendix_not_choi}
\end{equation}
The matrix $\rho_X$ is the \emph{normalised Choi state} of the
NOT channel. The active bit flip has been encoded in a static
bipartite state. For comparison, the identity channel leaves
$\ket{\Phi^+}$ unchanged: its Choi state is
$\rho_{\rm id}=\ket{\Phi^+}\!\bra{\Phi^+}$.

The encoding can be checked by an elementary correlation
measurement. Let $Z=\ket0\bra0-\ket1\bra1$. For the identity
channel, a $Z$ measurement before and after evolution returns the
same sign, whereas for the NOT channel it returns the opposite
sign. The same statements appear as bipartite correlations:
\begin{equation}
 \begin{aligned}
  \Tr[(Z\otimes Z)\rho_{\rm id}]&=+1,\\
  \Tr[(Z\otimes Z)\rho_X]&=-1.
 \end{aligned}
 \label{eq:appendix_not_correlation}
\end{equation}
These are also the sequential $Z$--$Z$ correlations when the
initial qubit is maximally mixed and the two measurements are
ideal projective measurements.

\subsection{Adding noise: an example closer to real experiments}
\label{app:noisy_bit_flip}

Suppose the NOT operation occurs with probability $p$, while with
probability $1-p$ nothing happens. The bit-flip channel is
\begin{equation}
 \mathcal E_p(\rho)=(1-p)\rho+pX\rho X.
 \label{eq:appendix_bitflip_channel}
\end{equation}
Its Choi state is simply the corresponding mixture,
\begin{equation}
 \rho_{\mathcal E_p}
 =(1-p)\ket{\Phi^+}\!\bra{\Phi^+}
 +p\ket{\Psi^+}\!\bra{\Psi^+},
 \label{eq:appendix_bitflip_choi}
\end{equation}
and
\begin{equation}
 \Tr[(Z\otimes Z)\rho_{\mathcal E_p}]
 =(1-p)(+1)+p(-1)=1-2p.
 \label{eq:appendix_bitflip_corr}
\end{equation}
Exactly the same $1-2p$ is obtained by measuring $Z$, applying the
noisy channel, and measuring $Z$ again. Thus a noisy \emph{process}
and a mixed bipartite \emph{state} encode the same specified
correlation. In the special cases $p=0$ and $p=1$, the Choi states
are pure; for $0<p<1$ the Choi state is mixed (of rank two).

\subsection{The general channel--state correspondence}
\label{app:choi}

The example suggests a general construction. Let $\mathcal H$
be $d$-dimensional, with an orthonormal basis
$\{\ket i\}_{i=1}^{d}$. Prepare a maximally entangled state
of an input qudit and a reference qudit,
\begin{equation}
 \ket{\Phi_d}=\frac1{\sqrt d}\sum_{i=1}^d\ket i\otimes\ket i.
 \label{eq:appendix_phi_d}
\end{equation}
Apply a quantum channel $\mathcal E$ to the first subsystem,
leaving the reference untouched. The resulting \emph{normalised}
Choi state is
\begin{equation}
 \rho_{\mathcal E}=
 (\mathcal E\otimes\mathrm{id})(\ket{\Phi_d}\bra{\Phi_d}).
 \label{eq:appendix_normalized_choi_definition}
\end{equation}
For a completely positive, trace-preserving channel,
$\rho_{\mathcal E}\geq0$ and $\Tr\rho_{\mathcal E}=1$;
its reference marginal is $\mathbb I/d$. In particular, a
unitary channel gives a pure Choi state, while channel noise is
represented by an appropriate density operator. This state
encodes the linear action of the channel on arbitrary inputs.

For calculations it is convenient to introduce the unnormalised
vector and Choi operator
\begin{equation}
 \ket\Omega=\sum_i\ket i\otimes\ket i,
 \qquad
 J_{\mathcal E}=(\mathcal E\otimes\mathrm{id})(\ket\Omega\bra\Omega)
 =d\rho_{\mathcal E}.
 \label{eq:appendix_cup}
\end{equation}
With the \emph{output} of $\mathcal E$ first and the untouched
reference second,
\begin{equation}
 J_{\mathcal E}=\sum_{i,j}
 \mathcal E(\ket i\bra j)\otimes\ket i\bra j.
 \label{eq:appendix_choi}
\end{equation}
For arbitrary operators $A$ and $B$ on $\mathcal H$,
\begin{equation}
 \boxed{\Tr\!\left[(B\otimes A^{\mathsf T})J_{\mathcal E}\right]
 =\Tr[B\mathcal E(A)]}. 
 \label{eq:appendix_choi_identity}
\end{equation}
Here $\mathsf T$ denotes transposition in the basis used in
Eq.~\eqref{eq:appendix_phi_d}. One can verify the identity
without diagrams: expand $A=\sum_{i,j}A_{ij}\ket i\bra j$,
then
\begin{align}
 \Tr[(B\otimes A^{\mathsf T})J_{\mathcal E}]
 &=\sum_{i,j}\Tr[B\mathcal E(\ket i\bra j)]
                \Tr[A^{\mathsf T}\ket i\bra j]\\
 &=\sum_{i,j}A_{ij}\Tr[B\mathcal E(\ket i\bra j)]
 =\Tr[B\mathcal E(A)].
 \label{eq:appendix_choi_proof}
\end{align}
Dividing by $d$ gives the corresponding identity for the
normalised Choi state,
\begin{equation}
 \Tr[(B\otimes A^{\mathsf T})\rho_{\mathcal E}]
 =\frac1d\Tr[B\mathcal E(A)].
 \label{eq:appendix_choi_normalized}
\end{equation}
Equation~\eqref{eq:appendix_bitflip_corr} is a simple instance
of this general formula.

\subsection{How the identity becomes a temporal measurement correlation}
\label{app:operational}

The trace identity is algebraic; its operational meaning can
also be derived directly. Prepare the maximally mixed input
$\rho_*=\mathbb I/d$. At the first time measure the dichotomic
projective observable $A=\Pi_+^A-\Pi_-^A$, allow the system
to evolve through $\mathcal E$, and then measure
$B=\Pi_+^B-\Pi_-^B$. Using the L\"uders update at the first
measurement, the outcome-weighted correlation is
\begin{align}
 C_{A,B}^{\rm temp}(\mathcal E)
 &=\sum_{a,b=\pm1}ab\,
   \Tr[\Pi_b^B\mathcal E(\Pi_a^A\rho_*\Pi_a^A)]\\
 &=\frac1d\Tr[B\mathcal E(A)]
 =\Tr[(B\otimes A^{\mathsf T})\rho_{\mathcal E}].
 \label{eq:appendix_temporal}
\end{align}
The second line follows because
$\sum_a a\Pi_a^A\rho_*\Pi_a^A=A/d$.
The direct physical interpretation is straightforward: the
first measured sign is imprinted on the input to the channel,
and the second measured sign is read at its output. The Choi
state records the same correlation between two tensor factors.

If two possible first observables $A_0,A_1$ and two possible
second observables $B_0,B_1$ are used with the \emph{same}
channel $\mathcal E$, then
\begin{equation}
 C_{xy}^{\rm temp}=\Tr[(B_y\otimes A_x^{\mathsf T})
 \rho_{\mathcal E}],\qquad x,y\in\{0,1\}.
 \label{eq:appendix_correspondence}
\end{equation}
The resulting temporal CHSH expression can therefore be
written as a spatial CHSH expression evaluated on one Choi
state. Fritz proved a broader equality of attainable quantum
correlation sets for the specified temporal and spatial CHSH
scenarios with projective measurements \cite{Fritz2010}.

For a general initial density operator $\rho$, the signed
postmeasurement state becomes
$\sum_a a\Pi_a^A\rho\Pi_a^A=\{A,\rho\}/2$;
thus the simple Eq.~\eqref{eq:appendix_temporal} corresponds
to the stated maximally mixed preparation and projective
instrument.

\subsection{Why the diagrams work: cup--cap duality and wire bending}
\label{app:wirebending}

The same calculation admits a compact diagrammatic form.
Write the unnormalised maximally entangled vector as
$\ket\Omega=\sum_i\ket{ii}$ and its dual as
$\bra\Omega=\sum_i\bra{ii}$. In tensor-network notation
$\ket\Omega$ is a \emph{cup}, and $\bra\Omega$ a \emph{cap}.
Their contractions obey the snake (or yanking) identity
\begin{equation}
 (\bra\Omega\otimes\mathbb I)(\mathbb I\otimes\ket\Omega)
 =\mathbb I.
 \label{eq:appendix_snake}
\end{equation}
Thus an input index can be bent around the cup and treated as
a second, reference-system index. The transpose in
Eq.~\eqref{eq:appendix_choi_identity} records the reversal of
its index orientation. This is the familiar ``wire-bending''
rule of tensor diagrams \cite{Penrose1971,CoeckeKissinger2017}.

\begin{figure}[htbp]
\centering
\begin{tikzpicture}[line width=0.85pt,font=\small]
  \draw (0,0) to[out=-90,in=-90] (1.8,0);
  \draw (0,0) -- (0,0.52);
  \node[draw,rounded corners,minimum width=0.95cm,minimum height=0.56cm]
     (chan) at (0,0.85) {$\mathcal E$};
  \draw (0,1.13) -- (0,1.55);
  \node[draw,minimum width=0.84cm,minimum height=0.52cm]
     (Bout) at (0,1.86) {$B$};
  \draw (1.8,0) -- (1.8,1.55);
  \node[draw,minimum width=0.95cm,minimum height=0.52cm]
     (Aref) at (1.8,1.86) {$A^{\mathsf T}$};
  \node[font=\footnotesize] at (0.9,-0.62) {Choi contraction};
  \node at (3.0,1.07) {$\Longleftrightarrow$};
  \node[draw,minimum width=0.84cm,minimum height=0.52cm]
     (Ain) at (4.25,0.03) {$A$};
  \draw (4.25,0.31) -- (4.25,0.58);
  \node[draw,rounded corners,minimum width=0.95cm,minimum height=0.56cm]
     (chan2) at (4.25,0.87) {$\mathcal E$};
  \draw (4.25,1.15) -- (4.25,1.57);
  \node[draw,minimum width=0.84cm,minimum height=0.52cm]
     (Bout2) at (4.25,1.86) {$B$};
  \node[font=\footnotesize] at (4.25,-0.62)
     {Input--channel--output};
\end{tikzpicture}
\caption{Wire bending translates a bipartite Choi contraction
(left) into an input--channel--output trace (right).
The ordering and the transpose are fixed by
Eq.~\eqref{eq:appendix_choi_identity}; the picture is a
representation of the algebraic identity.}
\label{fig:appendix_wirebend}
\end{figure}

\subsection{Return to the neuronal four-time LG experiment}
\label{app:ei-application}

For the collective E--I Hamiltonian $H_S=\heff\Omega\sigma_x/2$
and $Q=\sigma_z$, the stipulated ideal projective pairwise
measurements give
$C_{ij}=\cos[\Omega(t_j-t_i)]$.
The four-time combination at equal lag $\tau$ is therefore
\begin{equation}
 K_4(\tau)=3\cos(\Omega\tau)-\cos(3\Omega\tau),
 \qquad K_4^{\max}=2\sqrt2\quad
 \text{at }\Omega\tau=\frac\pi4.
 \label{eq:appendix_ei_fourtime}
\end{equation}
Its maximum equals the spatial Tsirelson bound.
The fixed-observable, four-time LG protocol uses the
four distinct pairs $(t_1,t_2)$, $(t_2,t_3)$,
$(t_3,t_4)$ and $(t_1,t_4)$; the last corresponds to
three times the nearest-neighbour evolution interval.
Hence these four measurements generally involve different
channels $\mathcal E_{ij}$ and are not automatically four
measurement settings on one fixed Choi state. The common
algebraic bound, the ideal quantum maximum, and the
channel--state correspondence together explain the structural
relationship between the two inequality families while
preserving the temporal character of our neuronal proposal.

\end{document}